\documentclass[aps,noshowpacs,twocolumn,superscriptaddress]{revtex4-1}  
\usepackage{graphicx}  
\usepackage{dcolumn}   
\usepackage{bm}        
\usepackage{amssymb}   
\usepackage{amsmath}
\usepackage{amsfonts}
\usepackage{bibunits}
\usepackage{braket}
\usepackage{gensymb}
\usepackage{siunitx}

\defaultbibliography{flatten_ws2_polariton}
\defaultbibliographystyle{ieeetr}

\begin{document}



\title{Room-temperature exciton-polaritons with two-dimensional WS$_2$}

\author{L. C.~Flatten} \email{lucas.flatten@materials.ox.ac.uk} \affiliation{Department of Materials, University of Oxford, Parks Road, Oxford OX1 3PH, United Kingdom} 
\author{Z.~He} \affiliation{Department of Materials, University of Oxford, Parks Road, Oxford OX1 3PH, United Kingdom}
\author{D. M.~Coles} \affiliation{Department of Materials, University of Oxford, Parks Road, Oxford OX1 3PH, United Kingdom} \affiliation{Clarendon Laboratory, Department of Physics, University of Oxford, OX1 3PU, United Kingdom}
\author{A. A. P.~Trichet} \affiliation{Department of Materials, University of Oxford, Parks Road, Oxford OX1 3PH, United Kingdom} 
\author{A. W.~Powell} \affiliation{Department of Materials, University of Oxford, Parks Road, Oxford OX1 3PH, United Kingdom} 
\author{R. A.~Taylor} \affiliation{Clarendon Laboratory, Department of Physics, University of Oxford, OX1 3PU, United Kingdom}
\author{J. H.~Warner} \affiliation{Department of Materials, University of Oxford, Parks Road, Oxford OX1 3PH, United Kingdom}
\author{J. M.~Smith} \affiliation{Department of Materials, University of Oxford, Parks Road, Oxford OX1 3PH, United Kingdom}
\vskip 0.25cm
\date{\today}

\begin{bibunit}

\begin{abstract}
Two-dimensional transition metal dichalcogenides exhibit strong optical transitions with significant potential for optoelectronic devices. In particular they are suited for cavity quantum electrodynamics in which strong coupling leads to polariton formation as a root to realisation of inversionless lasing, polariton condensation and superfluidity. 
Demonstrations of such strongly correlated phenomena to date have often relied on cryogenic temperatures, high excitation densities and were frequently impaired by strong material disorder. At room-temperature, experiments approaching the strong coupling regime with transition metal dichalcogenides have been reported, but well resolved exciton-polaritons have yet to be achieved. Here we report a study of monolayer WS$_2$ coupled to an open Fabry-Perot cavity at room-temperature, in which polariton eigenstates are unambiguously displayed. In-situ tunability of the cavity length results in a maximal Rabi splitting of $\hbar \Omega_{\rm{Rabi}} = $~\SI{70}{\milli\electronvolt}, exceeding the exciton linewidth.
Our data are well described by a transfer matrix model appropriate for the large linewidth regime. This work provides a platform towards observing strongly correlated polariton phenomena in compact photonic devices for ambient temperature applications. 
\end{abstract}

\pacs{}
\maketitle
\begin{figure*}
\includegraphics[width=0.55\textwidth]{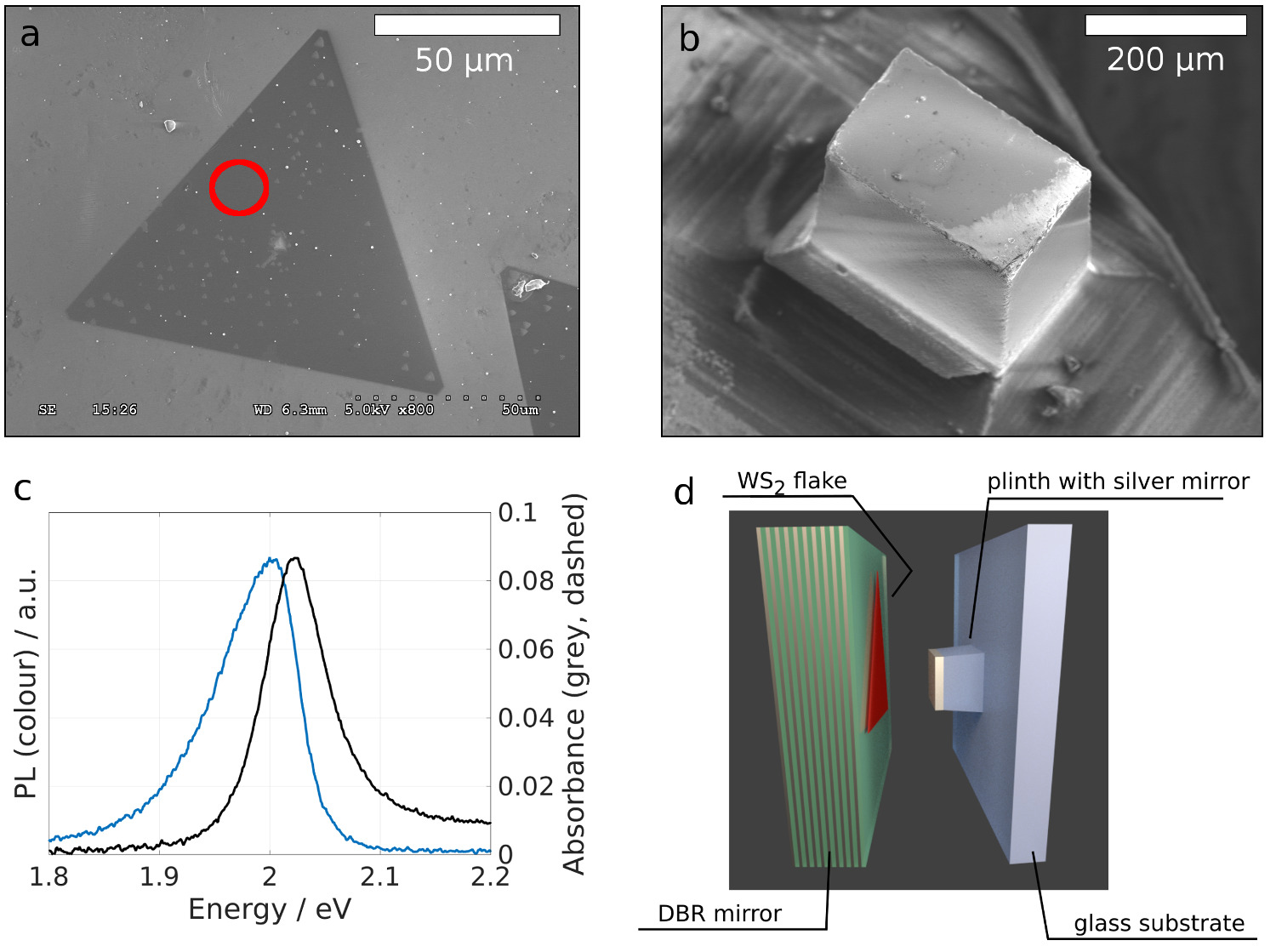}
\caption[justified]{\textbf{WS$_2$ in an open microcavity.} a) SEM micrograph of a triangular monolayer WS$_2$ flake transferred onto a DBR mirror. b) Elevated silica plinth with a silver coating forming one side of the microcavity. c) Absorbance (black, dashed) and photoluminescence (colour) spectra after off-resonant, continuous wave excitation ($\lambda_{\rm{exc}} = $ \SI{473}{\nano\meter}) of marked region in a. d) Schematics of cavity setup: The left side consists of a DBR with 10 pairs of SiO$_2$/TiO$_2$ with single WS$_2$ flakes transferred to the low refractive-index terminated surface. The cavity is formed by positioning a silver mirror opposite the DBR.  \label{fig1}}
\end{figure*}

Transition metal dichalcogenides (TMDCs) have received increased attention due to the ability to produce large, atomically flat monolayer domains with intriguing optical properties \cite{ gutierrez_extraordinary_2013, rong_controlling_2014, jiang_photonic_2014, chen_q-switched_2015, vasilevskiy_exciton_2015,lu_interactions_2016}. Their large exciton-binding energy leads to stable exciton formation at room temperature, narrow absorption peaks and high photoluminescence quantum yields \cite{zhu_exciton_2015, zhao_evolution_2013,xu_spin_2014,scrace_magnetoluminescence_2015,amani_near-unity_2015}.
Recently it has become possible to grow atomically thin single-crystal domains of WS$_2$ by chemical vapour deposition (CVD) \cite{rong_controlling_2014}. WS$_2$, like MoSe$_2$ and WSe$_2$ transits from being an indirect bandgap material in bulk to having a direct bandgap as a monolayer. Owing to this direct bandgap WS$_2$ interacts strongly with light: even though it is a single atomic layer with a thickness of \SI{0.8}{\nano\meter} absorbance values of 0.1 and strong photoluminescence (PL) can be observed (see Fig. \ref{fig1}c) \cite{zhao_evolution_2013}. The valley polarisation degree of freedom \cite{yu_valley_2015} and a finite Berry curvature \cite{srivastava_signatures_2015} make the material suitable for spinoptronics. In particular the large exciton binding energy of $\approx 0.7$ eV makes monolayer WS$_2$ suitable for room temperature applications \cite{zhu_exciton_2015}, enabling polariton formation at such high temperatures.

Incorporated into a microcavity with sufficiently small mode volume the two-dimensionally confined exciton couples to the photonic modes resulting in the formation of new eigenstates of the system, the polariton states, that consist of an admixture of the uncoupled states. 
Polaritons retain the characteristics of their constituent parts e.g. they have a degree of delocalization gained from the photon, while retaining a mass (typically $\sim 10^{-4}$ that of the free electron mass \cite{kasprzak_boseeinstein_2006}) and a finite scattering cross section inherited from the exciton, which gives rise to non-linear effects \cite{
dang_stimulation_1998,
senellart_nonlinear_1999,
bhattacharya_room_2014,
schneider_electrically_2013} and strongly correlated phenomena \cite{kasprzak_boseeinstein_2006,
amo_superfluidity_2009, byrnes_exciton-polariton_2014, plumhof_room-temperature_2014}.

In this Letter we present the first study in which WS$_2$ is introduced experimentally to an open microcavity to take advantage of its extraordinary optical properties for exciton-polariton formation. We make use of a cavity setup which enables in-situ tunability of the coupling strength between the optical mode and WS$_2$ excitons. Strong coupling between a photonic mode and MoS$_2$ has been reported in a monolithic microcavity \cite{liu_strong_2015}, but suffered from poorly resolved spectral features with splittings below the exciton linewidth. At low temperatures progress with transversely confined microcavities have been made \cite{schwarz_two-dimensional_2014,dufferwiel_exciton-polaritons_2015}, allowing polariton formation in MoSe$_2$. The approach we present here entails unambiguous room temperature polariton formation with a Rabi splitting of $\hbar \Omega_{\rm{Rabi}} = $~\SI{70}{\milli\electronvolt}, exceeding the exciton linewidth and allowing in-situ variability of the coupling strength. We extend the theoretical description of classical polariton formation to the room-temperature large linewidth regime, for which we found corrections to current theory \cite{savona_quantum_1995}. Furthermore we investigate the polariton distribution as a function of cavity detuning and coupling strength. Our findings establish a platform towards integrated polariton devices at room temperature suitable for spinoptronics and strongly correlated phenomena.

\section*{\label{results} Results}
The CVD grown WS$_2$ flakes have lateral dimensions exceeding \SI{100}{\micro\meter} (Fig. \ref{fig1}a), which are transfered with a PMMA transfer layer onto a low-index terminated distributed Bragg reflector (DBR). The opposite side of the microcavity is formed by a small silver mirror (Fig. \ref{fig1}b), which has a lower reflectivity than the DBR resulting in a cavity finesse of $F \approx 50$. By placing the silver mirror opposite a region of the DBR holding WS$_2$ (Fig. \ref{fig1}d) and varying the cavity length with a Piezo microactuator, stable cavity modes interacting strongly with the WS$_2$ can be obtained (see Methods).
\begin{figure*}
\includegraphics[width=1\textwidth]{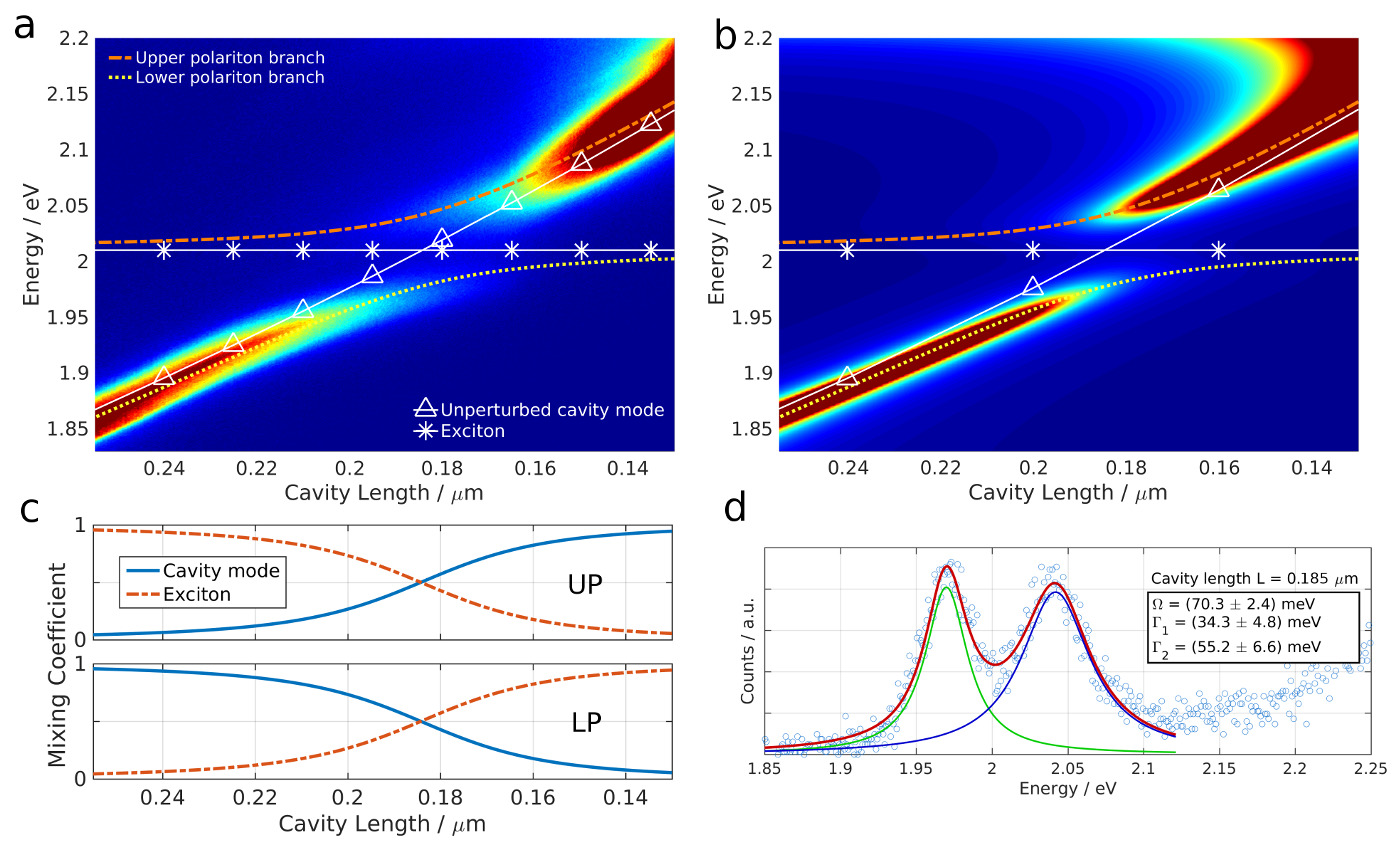}
\caption[flushleft]{\textbf{Polariton dispersion at room temperature.} a) Transmission spectra as the cavity length is swept, showing typical polariton dispersion. The white continuous lines show the energy of the unperturbed cavity mode (triangles) and the exciton (stars). The energy of the coupled system, the two polariton branches are shown with the coloured, dashed lines. b) Transmission spectra as cavity length is swept obtained by transfer matrix modelling. The dispersion lines from a) are overlaid. c) Photonic and excitonic fractions of upper (UP) and lower (LP) polariton branch. d) Transmission spectrum at maximal photon-exciton mixing revealing a Rabi Splitting of $\Omega_{\rm{Rabi}} = 70$ meV. \label{fig2}}
\end{figure*}

Fig. \ref{fig2}a shows successively acquired transmission spectra for different cavity lengths tracking the mode with longitudinal mode number $q = 3$. The cavity length is decreased from left to right from  \SI{260}{\nano\meter} to \SI{130}{\nano\meter} which leads to a linear response in cavity mode energy moving from \SI{1.85}{\electronvolt} to \SI{2.15}{\electronvolt}. The exciton energy of the monolayer  WS$_2$ stays constant at \SI{2.01}{\electronvolt} (white lines in Fig. \ref{fig2}a). These two states are the uncoupled photon and exciton states, which couple in our system and show a typical avoided level crossing, forming the upper (UP) and lower polariton (LP) branch (coloured and dashed lines). Fig. \ref{fig2}b shows the calculated transmission spectra obtained with a transfer matrix method (TMM). The strongly coupled system can be described by the Hamiltonian
\begin{equation}
\begin{split}
H = \  & E_{\rm{cav}} b^\dagger b
+ E_{\rm{exc}} x^\dagger x +  V (b^\dagger x + c.c. )
\end{split}
\label{eq2}
\end{equation}
where $E_{\rm{cav}}$ and $E_{\rm{exc}}$ correspond to the energy levels of cavity mode and exciton which are coupled by the interaction potential $V = \frac{\Omega_{\rm{Rabi}}}{2}$. $b^\dagger,b$ and $x^\dagger,x$ are the respective creation and annihilation operators. The system can be reduced to:
  \begin{equation}
H_{\rm{Int}} \begin{pmatrix}\alpha \\ \beta \end{pmatrix}
=
\begin{pmatrix} 
E_{\rm{cav}} & \frac{\Omega_{\rm{Rabi}}}{2} \\
\frac{\Omega_{\rm{Rabi}}}{2} & E_{\rm{exc}} 
\end{pmatrix}\begin{pmatrix} 
\alpha \\
\beta
\end{pmatrix}  = E \begin{pmatrix} \alpha \\ \beta \end{pmatrix}
\label{eq1}
\end{equation}
The eigenstates of the equation represent superpositions of the bare states, photonic mode and exciton, which are called polaritons. The coefficients $\alpha^2$ and $\beta^2$ quantify the contribution of photonic and excitonic part respectively, and are plotted in Fig. \ref{fig2}c. As the cavity mode is tuned through the exciton energy the lower (upper) polariton branch switches from photon- (exciton)-like to exciton- (photon)-like. Fig. \ref{fig2}d contains a vertical slice through the data presented in Fig. \ref{fig2}a at the crossing point of exciton and photon energy. The Rabi splitting is evaluated to $\hbar \Omega_{\rm{Rabi}} = (70 \pm 2)$~\SI{}{\milli\electronvolt} from a fit to the data using two Lorentzian lineshapes. The splitting is fully resolved, since the individual linewidths of upper and lower polariton branch are $(55 \pm 7)$ and $(34 \pm 5)$~\SI{}{\milli\electronvolt} respectively, smaller than $\hbar \Omega_{\rm{Rabi}}$. Note that the cavity mode linewidth increases from $\approx 30 $~\SI{}{\milli\electronvolt} to $\approx 60 $~\SI{}{\milli\electronvolt} across the energy window presented in Fig. \ref{fig2} even without the presence of an absorber due to the edge of the stop-band of the DBR which is centered around \SI{1.95}{\electronvolt} ($\lambda = $~\SI{637}{\nano\meter}). The agreement between TMM data and the experimentally obtained spectra is excellent.

\begin{figure*}
\includegraphics[width=0.55\textwidth]{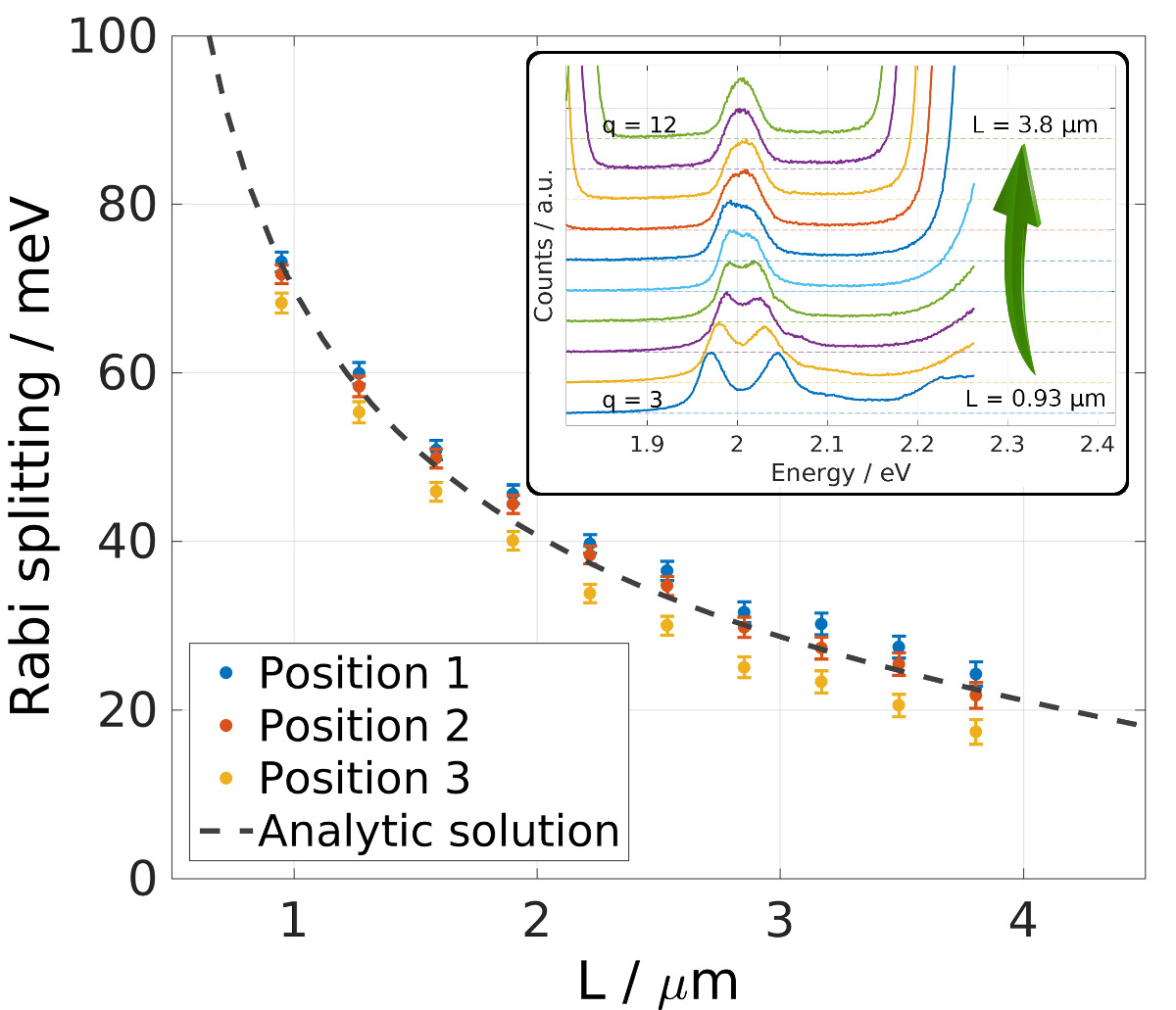}
\caption[flushleft]{\label{fig3}\textbf{Varying the Rabi splitting.} Rabi Splitting for different longitudinal cavity modes ($q = 3,...,12$). Symbols show experimentally obtained values for three different positions on the WS$_2$ flake with corresponding uncertainty. The dashed line corresponds to the analytic expression shown in Eq. \ref{eqSolv}, derived in the Suppl. Materials. The inset depicts transmission spectra for the different longitudinal mode indices stacked vertically for better visualisation.}
\end{figure*}

The open-access design of the microcavity allows for opening the cavity freely, which gives access to different longitudinal mode numbers $q$. For increasing cavity length (increasing $q$) the coupling strength between cavity mode and the WS$_2$ monolayer decreases. Fig. \ref{fig3} presents data obtained by evaluating the polariton dispersion for the first ten accessible longitudinal mode numbers ($q = 3,...,12$). The two polariton branches are described by two Lorentzian peaks, whose peak positions are fitted with the dispersion obtained analytically from diagonalisation of Eq. \ref{eq1}. The Rabi splitting $\hbar \Omega_{\rm{Rabi}}$ is obtained as one parameter of this fit and plotted as symbols in Fig. \ref{fig3}. The inset in Fig.~\ref{fig3} displays the transmission spectra at the crossing point for the different longitudinal mode indices. The dashed line shows the analytic solution for the Rabi splitting derived by solving Maxwell's equations for the specific cavity geometry. Thus:
\begin{equation}
\begin{split}
\hbar \Omega(L) = & 2 \hbar \sqrt{ V^2  - 
     \frac{1}{4}(\gamma_x - \gamma_p)^2)}
     \label{eqSolv}
     \end{split}
\end{equation}
\begin{equation}
\begin{split}
\gamma_p  =  & \frac{1-\sqrt{R}}{\sqrt{R}} \frac{c}{2 n_c L_{\rm{eff}}}  \\ \
V^2  =  & \frac{1 + \sqrt{R}}{\sqrt{R}} \frac{k d W c}{2 n_c^2 L_{\rm{eff}} } + \frac{k d W}{2 n_c} \left(\gamma_p + \gamma_x \right)   + \left(\frac{k d \epsilon_B c}{2L n_c^2 \sqrt{R}}\right)^2
\end{split}
\label{eq22}
\end{equation} 
Here $R$ is the mirror reflectivity (in our case $R_{\rm{Silver}}\ll 
R_{\rm{DBR}}$, so that we can set $R = R_{\rm{Silver}}$), $L_{eff} = L + L_{\rm{DBR}}$ 
($L$ is the geometric cavity length, $L_{\rm{DBR}}$ is the effective length of the DBR \cite{kavokin_microcavities_2011}), $\gamma_p, \gamma_x$ are the cavity and exciton half-widths (HWHM), $c$ the speed of light, $n_c$ the refractive index within the cavity, $d$ the width of the monolayer ($d = 0.8$~\SI{}{\nano\meter}), $k = \frac{E_{\rm{exc}}}{\hbar c}$, $\epsilon_B$ the dielectric background and $W$ is a parameter proportional to the oscillator strength of the WS$_2$ monolayer. We derive these expressions in the supplementary material. $\epsilon_B$ and $W$ can be directly obtained from the dielectric function of monolayer WS$_{\textrm{2}}$ \cite{li_measurement_2014}, thus the system has no free parameter. For the fit to the data presented in Fig. \ref{fig3} it is sufficient to measure the absorbance of the flake, which shows slight spatial variations for our sample, and obtain $W$ in this way (see Fig. \ref{fig3}). Note that Eq. \ref{eq22} has been published before for quantum wells without the second and the third term in the expression for $V^2$ \cite{kavokin_microcavities_2011,savona_quantum_1995}. The first correction term is small for $\gamma_p + \gamma_x \ll \frac{2 c}{n_c L}$, but for room temperature polariton applications it becomes sizable. In our case it corrects the value for $V^2$ by $7\%$ for $L = 1$~\SI{}{\micro\meter} and by $25\%$ for $L = 4$~\SI{}{\micro\meter}. The second correction term can be neglected if $\frac{k_d d \epsilon_B c}{4 L n_c^2} \ll W$, which in our system is only the case for cavity lengths exceeding $L \approx 2.5$~\SI{}{\micro\meter} where it contributes less than $10\%$ (see Suppl. Information for a detailed discussion).
\begin{figure*}
\includegraphics[width=1.0\textwidth]{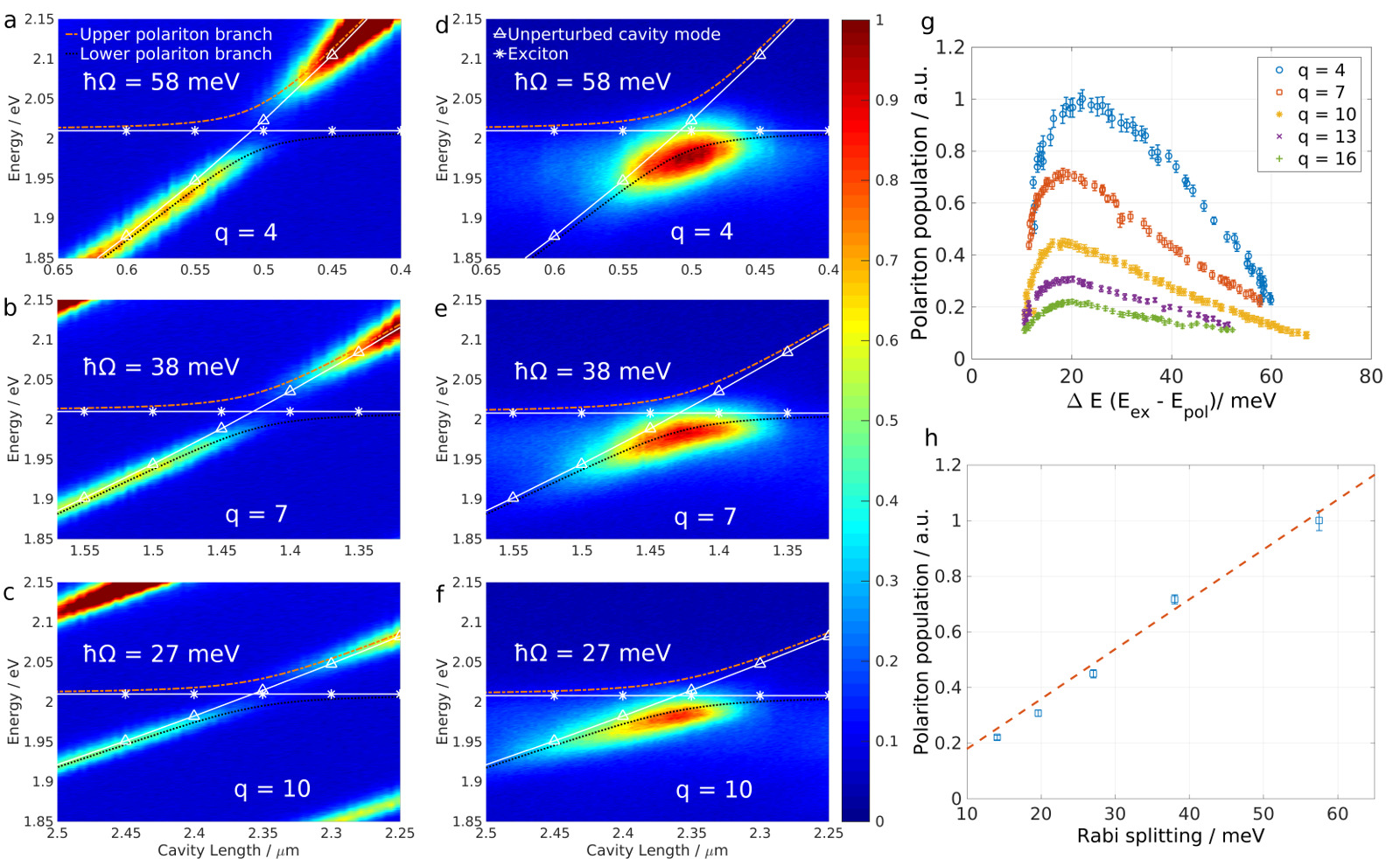}
\caption[flushleft]{\textbf{Polariton population with off-resonant pump.} Transmission (a-c) and PL (d-f) spectra as cavity length is swept, traversing different longitudinal modes ($q =$~4, 7, 10). For PL measurements the sample is excited off-resonantly with a ($\lambda_{\rm{exc}} = $ \SI{473}{\nano\meter}) continuous-wave laser. The lines show the dispersion of the uncoupled (white) and coupled system (colour) as in Fig. \ref{fig1}. g) Polariton population as obtained from Lorentzian lineshape fits to the PL data for $q = 4, 7, 10, 13$ and 16. h) Maximal polariton population as a function of the Rabi splitting $\hbar\Omega$. Dashed line shows a linear dependence to guide the eye. \label{fig4}}
\end{figure*}
Interestingly the change in cavity length and the subsequent modulation of the Rabi splitting has stark consequences for polariton dynamics. When performing an optical transmission experiment polariton states are probed by coupling to the photonic component, which leads to the typical changes in intensity across a branch. A different way to populate exciton-polariton states is to optically pump the excitons non-resonantly, causing scattering and direct radiative pumping. Figs. \ref{fig4}a-c display white lamp transmission data for higher $q$ values $q = 4, 7$ and $10$, showing a decreasing degree of coupling. The overlaid lines depict the uncoupled (white, continuous) and coupled (colour, dashed) dispersions as before (comp. Fig. \ref{fig2}). For the datasets presented in Figs. \ref{fig4}d-f the cavity length is scanned in the same way as for the transmission datasets, but here the sample is excited with a $\lambda = $ \SI{473}{\nano\meter} continuous wave laser. The resulting spectra show that only the central region of the lower polariton branch is populated. In particular the intensity and shape of the emission along the branch varies for different longitudinal mode numbers for the same laser irradiation. It is thus a function of the Rabi splitting. Note that the modulation of the incident laser power due to the change in cavity length is negligible, as the pump wavelength is far below the stopband of the mirror through which the monolayer is excited. Fig. \ref{fig4}g shows the polariton population for $q = 4, 7, 10, 13$ and $16$ as a function of the energy difference $\Delta E$ between polariton branch and exciton energy ($\Delta E = E_{\rm{exc}} - E_{\rm{LP}})$. It is obtained by fitting a Lorentzian lineshape to the PL data presented in Figs. \ref{fig4}d-f and scaling the obtained amplitudes by the inverse of the square of the photonic fraction $\alpha$ for the respective polariton branch. In general the lower polariton branch is populated slowly while its energy approaches the exciton energy from below. The population reaches a maximum between $\SI{15}{\milli\electronvolt} < \Delta E <  \SI{30}{\milli\electronvolt}$ and decreases rapidly for $\Delta E \rightarrow 0$. The absolute polariton population number is not the same for different longitudinal cavity modes and decreases for a smaller Rabi splitting. In fact Fig. \ref{fig4}h shows the maximum polariton population as a function of the associated Rabi splitting, revealing a linear dependence. This trend could be explained by the scattering rate from exciton reservoir to polariton state, which is approximately proportional to the energy difference between the two states \cite{michetti_strongly_2015} and governs the polariton population in the steady state. The upper polariton branch stays unpopulated for all cases.

\section*{\label{conclusion} Discussion}
The off-resonant pump leads to excitation high in the conduction band of the WS$_2$ monolayer which is followed by a rapid thermalisation, creating an exciton bath which then populates the lower polariton branch \cite{coles_vibrationally_2011,
coles_polariton-mediated_2014}. Two known pathways for populating processes are the direct radiative decay channel, whose rate is proportional to the photonic coefficient $\alpha$ of the LPB and the phonon-assisted scattering of excitons into the LPB which is proportional to the excitonic coefficient $\beta$ \cite{michetti_exciton-phonon_2009}. On the other hand there are multiple relaxation pathways for polaritons: the direct radiative decay proportional to $\alpha$ dominating for large $\Delta E$, exciton-electron scattering proportional to $\beta$ \cite{sidler_fermi_2016} and exciton-exciton annihilation \cite{yuan_exciton_2015} proportional to $\beta^2$. The processes proportional to powers of $\beta$ result in the fast decay of polaritons for small $\Delta E$. The quantitative description of this system bears potential for further studies.

In conclusion we have demonstrated strong coupling between photonic cavity modes and excitons in two-dimensional atomically-thin WS$_{\textrm{2}}$ at room temperature. The coherent exchange of energy between those two constituents results in the formation of exciton-polaritons with a vacuum Rabi splitting of $70 \pm \SI{3}{\milli\electronvolt}$. Monolayers of WS$_{\textrm{2}}$ represent a promising candidate for polariton based devices due to their large exciton binding energies allowing for room temperature operation and very large oscillator strengths. We demonstate in situ control over the coupling strength and derive an analytic expression describing the length dependence. By studying the PL of the device we observe that the coupling strength directly influences the population dynamics. 

Strongly coupled devices of the type described here could provide a route towards observing strongly correlated phenomena at room temperature. In particular properties such as the  valley polarisation degree of freedom \cite{yu_valley_2015} and a finite Berry curvature \cite{srivastava_signatures_2015} render devices based on TMDCs attractive for spinoptronics and quantum computation.

\section*{\label{setup} Methods}
\subsection*{Sample preparation}
The open microcavity consists of two opposing flat mirrors, a large dielectric DBR with 10 pairs of SiO$_2$, TiO$_2$ with central wavelength of $\lambda = $ \SI{637}{\nano\meter} and a smaller silver mirror with a thickness of $\SI{50}{\nano\meter}$, deposited via thermal evaporation (Fig. \ref{fig1}). 
The WS$_{\textrm{2}}$ flakes are transferred onto the dielectric mirror stack, which has a low refractive-index terminated configuration to provide an anti-node of the electric field at the mirror surface and thus optimal coupling to the monolayer.  Note that this condition rules out the recently described Tamm-plasmonic coupling, as it requires a high refractive-index termination before the metallic layer \cite{lundt_room_2016}. This transfer process is facilitated by coating the as-grown WS$_{\textrm{2}}$ flakes on SiO$_2$ with a helper layer of PMMA. After etching away the substrate, the floating PMMA film was transfered manually onto the DBR and baked at 150$^\circ$ for 15 min. The remaining PMMA was then dissolved by placing the sample in an acetone bath for 10 min. After non-resonant excitation with a $\lambda = \SI{473}{\nano\meter}$ laser the WS$_2$ flakes show strong neutral exciton emission at \SI{2.01}{\electronvolt} with little, spatially varying contribution from the charged exciton state at 1.98 eV (see Fig. \ref{fig1} c), which we attribute to a non-uniform excess electron background \cite{zhu_exciton_2015}.
The small silver mirror is mounted on a three-dimensional piezo actuated stage, which makes positioning of the silver mirror relative to the WS$_{\textrm{2}}$ flake possible and allows for electrical control of the cavity length. 
\subsection*{Optical measurements}
By positioning the silver plinth over a region of the DBR mirror which holds monolayer WS$_{\textrm{2}}$ and reducing the distance between the two mirrors below $\approx$~\SI{5}{\micro\meter}, stable cavity modes interacting strongly with the WS$_{\textrm{2}}$ excitons will appear in a transmission experiment. The light is focused onto an Andor combined spectrometer/CCD for analysis. The setup allows for off-resonant optical excitation below the stop-band of the DBR with a continuous wave laser with wavelength $\lambda = \SI{473}{\nano\meter}$ at power densities around $\rho_{\rm{exc}} = 1500 \frac{\rm{W}}{\rm{cm}^2}$.

\section*{Acknowledgements}
We thank Radka Chakalova at the Begbroke Science Park for helping with the thermal evaporation and dicing of the mirrors. L.F. and A.T. acknowledge funding from the Leverhulme Trust. D.C. acknowledges funding from the Oxford Martin School and EPSRC grant EP/K032518/1.

\section*{Author contributions}
Z.H. grew the WS$_{\textrm{2}}$ and transferred the flakes onto the mirror. L.F. prepared all other parts of the setup and conducted the experiments. L.F., D.C. and J.S. conceived the idea and prepared the manuscript. L.F., A.P., A.T. and J.S. worked on the analytical description. R.T., J.W. and J.S. oversaw the experiments and revised the manuscript.

\section*{Additional Information}
This paper is accompanied by supplementary information. Additionally the data presented in this work is available at https://ora.ox.ac.uk:443/objects/uuid:d3f0b229-df66-4895-ba1e-0d9ef9a07ec9. During the publication process, the authors became aware of a related study by Wang et al. \cite{wang_coherent_2016}, demonstrating exciton-polaritons with monolayer WS$_2$ incorporated into a monolithic metallic cavity. The authors declare no competing financial interests.
\putbib

\end{bibunit}

\clearpage
\pagebreak
\widetext
\appendix
\begin{center}
\textbf{\large Supplemental Materials: Room-temperature exciton-polaritons with two-dimensional WS$_{\textrm{2}}$}
\end{center}
\setcounter{equation}{0}
\setcounter{figure}{0}
\setcounter{table}{0}
\setcounter{page}{1}
\makeatletter
\renewcommand{\theequation}{S\arabic{equation}}
\renewcommand{\thefigure}{S\arabic{figure}}
\renewcommand{\bibnumfmt}[1]{[S#1]}
\renewcommand{\citenumfont}[1]{S#1}
\begin{bibunit}

\section{Details of experimental setup}
The DBR mirror consists of 10 pairs of SiO$_2$, TiO$_2$ with refractive indices 1.45 and 2.05 respectively and the stopband is centered around $\lambda = $ \SI{637}{\nano\meter}. The small mirror is produced by removing large areas of a flat silica substrate with a dicer to create a 200 $\times$ 300 \SI{}{\micro\meter}$^2$ plinth made semi-reflective by thermally evaporating a \SI{50}{\nano\meter} thick silver layer with a reflectivity of $R = 95\%$. It is mounted on a three-dimensional piezo-actuated stage, which makes electronic positioning of the silver mirror relative to the WS$_{\textrm{2}}$ flake possible. To initialise the cavity white light from a light emitting diode is shone through the mirrors while reducing the separation. With the help of Fabry-Perot fringes visible for small mirror separations ($L < $ \SI{20}{\micro\meter}) the substrates are made parallel within \SI{150}{\micro\radian}.
Optical access to the sample is given by a standard $\times 10$ objective lens and the collected light is focused on an Andor combined spectrograph/CCD with a 300 grooves/mm grating. For the photoluminescence experiment the sample is excited with a PicoQuant D-C-470 laser with $\lambda = \SI{473}{\nano\meter}$ at power densities around $\rho_{\rm{exc}} = 1500 \frac{\rm{W}}{\rm{cm}^2}$.

\section{Derivation of analytic expression for Rabi splitting}
\subsection{Transfer matrices}
Maxwell's equation in a spatially homogeneous medium with refractive index $n$ reduce to
\begin{equation}
\nabla^2 \vec{E} + k^2 n^2 \vec{E} = 0
\end{equation}
with the vacuum wavenumber $k = \frac{2 \pi}{\lambda}$. In particular we get $\partial_z^2 E = -k^2 n^2 E$ for the field amplitude in $z$-direction. A general form of the solution for waves in this direction is
\begin{equation}
E(z) = A^+ e^{i n k z} + A^- e^{-i n k z}
\label{eqS2}
\end{equation}
$A^+$ and $A^-$ are the amplitudes of forward and backward travelling waves. For the transition from a medium with refractive index $n_1$ to one with $n_2$ we can define the amplitude reflection and transmission coefficients:
\begin{equation}
\begin{split}
r = \frac{A_1^-}{A_1^+} \qquad \& \qquad t = \frac{A_2^+}{A_1^+}
\end{split}
\end{equation}
To solve the boundary conditions for multiple interfaces it is convenient to define a transfer matrix T, which propagates the wave across a homogeneous layer. The analytic form of such a matrix can be obtained after chosing a convenient basis set. One possibility is to define the vector \cite{kavokin_microcavities_2011}
\begin{equation}
\Phi(z) = \begin{pmatrix}
E(z) \\ \frac{1}{i k} \partial_z E(z)
\end{pmatrix}
\end{equation}
A matrix that enforces the propagation through a layer with refractive index $n$ and thickness $a$, $T_a \Phi | _{z = 0} = \Phi | _{z = a}$ has the form
\begin{equation}
T_a = \begin{pmatrix}
\cos( n k a) & \frac{i}{n}\sin( n k a) \\
in \sin( n k a) & \cos( n k a)
\end{pmatrix}
\end{equation}
\subsection{Application to dispersive medium in planar cavity}
\begin{figure}[h!]
\centering
\includegraphics[width=0.5\textwidth]{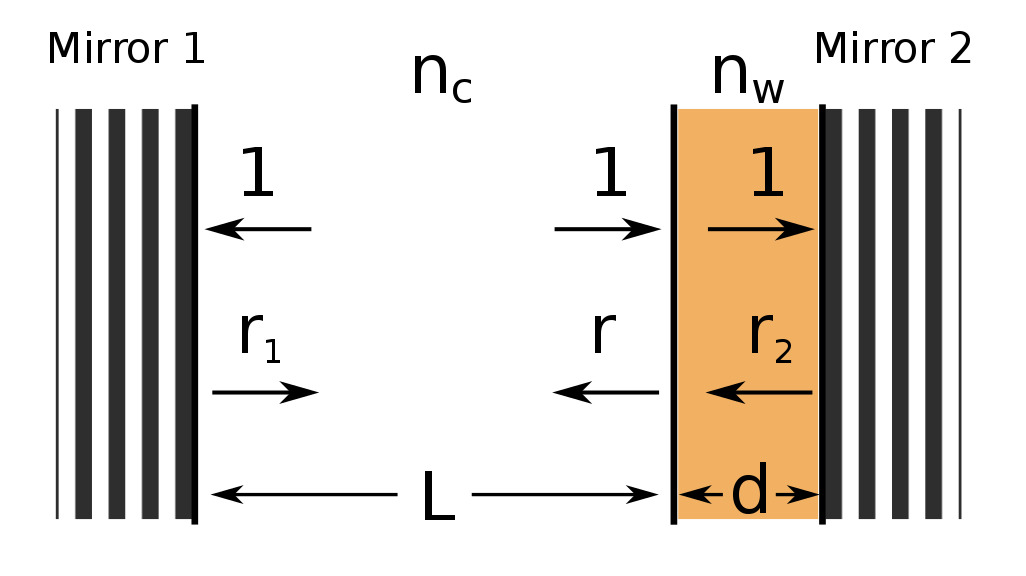}
\caption[flushleft]{\textbf{A thin dispersive medium inside a planar microcavity.} Schematics of planar cavity with mirror separation $L + d$ and refractive index $n_c$ for the cavity medium and $n_w$ for the thin dispersive medium. The arrows give the amplitudes of forwards and backwards propagating waves as introduced in the text. \label{figS1}}
\end{figure}
We reduce the problem of finding the resonant modes in the cavity to finding the amplitude reflection coefficient $r$ as defined in Fig. \ref{figS1}. Having found $r$ we can compute the cavity dispersion by solving ($k_c = n_c k$)
\begin{equation}
\begin{pmatrix}
\cos(k_c L) & \frac{i}{n_c}\sin(k_c L) \\
in_c \sin(k_c L) & \cos(k_c L)
\end{pmatrix}  \begin{pmatrix}
1+r_1 \\ n_c (r_1 -1)
\end{pmatrix} = A 
\begin{pmatrix}
1+r \\ n_c (1 - r) 
\end{pmatrix}
\end{equation}
The choice of coefficients for the vectors on each side of the cavity ensures the correct reflection conditions, i.e. attenuation and phase jump. Eliminating $A$ we obtain
\begin{equation}
r_1 r e^{i 2 k_c L } = 1
\label{eqS1}
\end{equation}
For ideal dielectric mirrors we would have $r_1 = r = 1$ and thus the well known Fabry-Perot resonance condition $k_c L = \pi q$ emerges, $q$ being an integer. Now the problem is reduced to finding $r$.
Taking the right side of the sketch in Fig. \ref{figS1} we can start with ($k_w = n_w k$)
\begin{equation}
\begin{pmatrix}
\cos(k_w d) & \frac{i}{n_w}\sin(k_w d) \\
i n_w \sin(k_w d) & \cos(k_w d)
\end{pmatrix}  \begin{pmatrix}
1+r \\ n_c (1 - r)
\end{pmatrix} = A 
\begin{pmatrix}
1+r_2 \\ n_w (1 - r_2) 
\end{pmatrix}
\end{equation}
Eliminating $A$ and solving for $r$ we get:
\begin{equation}
r = \frac{n_c - n_w + e^{i 2 k_w d}(n_c + n_w) r_2}{n_c + n_w + e^{i 2 k_w d}(n_c - n_w) r_2}
\end{equation}

This expression is further reducible if we assume a certain form for $r_2$. A suitable choice would be the reflection coefficient for a distributed Bragg reflector $r_{\rm{DBR}}$ or metallic mirror $r_{\rm{Met}}$ (derived in \cite{kavokin_microcavities_2011}):
\begin{equation}
r_{\rm{DBR}} = \sqrt{R} e^{i (k - k_{\rm{DBR}}) \frac{L_{\rm{DBR}}}{n_w}} \qquad \& \qquad r_{\rm{Met}} = \frac{n_w - n_m}{n_w + n_m}
\end{equation} 
Here $R$ is the mirror reflectivity, $L_{\rm{DBR}} = \frac{n_a n_b \lambda_{\rm{DBR}}}{2(n_b-n_a)}$ is the effective length of the DBR, with $\lambda_{\rm{DBR}}$ the central wavelength, $k_{\rm{DBR}} = \frac{2 \pi}{\lambda_{\rm{DBR}}}$ and $n_m$ the complex refractive index of the metal. If we assume a perfect dielectric reflector ($r_2 = 1$) we obtain:
\begin{equation}
r = \frac{n_c\cos(k_w d) + i n_w \sin(k_w d)}{n_c \cos(k_w d) - i n_w \sin(k_w d)}
\label{eqRef}
\end{equation}

Substituting this expression into Eq. \ref{eqS1}, assuming a thin layer ($|k_w d| << 1$) and setting $r_1 = 1$ 
we obtain:
\begin{equation}
n_c \sin(k_c L) + n_w k_w d \cos(k_c L) = 0
\end{equation}
Evaluating this expression close to the cavity mode resonance ($\cos(k_c L) \approx 1, \sin(k_c L) \approx n_c(k-k_p)L$, with $k_p$ being the wavenumber on resonance) we get to the succinct expression: 
\begin{equation}
\qquad n_c^2 (k-k_p) L + n_w^2 k d = 0
\label{eqMain}
\end{equation}
Now we want to substitute in $n_w$, the complex refractive index of the dispersive thin layer, and find the dispersion of the resonant modes. The dielectric function of a system of Lorentz oscillators as derived from solving the optical Bloch equations can be written as \cite{khitrova_nonlinear_1999}
\begin{equation}
\varepsilon(k) = \varepsilon_{\rm{B}} - \frac{W}{c (k - k_x) + i \gamma_x}
\end{equation}
where $W$ is proportional to the oscillator strength of the material, $k_x$ is the wavenumber of the oscillator resonance and $\gamma_x$ is the oscillator HWHM linewidth. The relation between $W$ and microscopic material parameters is derived in  \cite{loudon_quantum_2000} and reads 
\begin{equation}
W = \frac{N \mu^2}{3 \varepsilon_0 \hbar V}
\label{loudon}
\end{equation}
with $\frac{N}{V}$ the number of oscillators per volume and $\mu$ the transition dipole moment for a single oscillator. From now on we set $\hbar = c = 1$. Putting $n_w = \sqrt{\varepsilon(k)}$ into Eq. \ref{eqMain} we obtain:
\begin{equation}
n_c^2 (k- k_p + i \gamma_p) L + \left( \varepsilon_{\rm{B}} - \frac{W}{k - k_x + i \gamma_x} \right) k d = 0
\end{equation}

At this stage we have included a term $i \gamma_p$ for the finite linewidth of the cavity mode. Of course this is artifical at this point and we will derive $\gamma_p$ in terms of the reflectivity $R$ of the mirrors later. We assume that $k$ varies little over the extent of the thin dispersive layer, in other words the factor $k d$ is constant ($k d = k_d d$). We can then solve the equation for $k$ and find by neglecting higher orders in $d$:
\begin{equation}
k_{1,2} = \frac{1}{2} \left( k_p + k_x - i (\gamma_p + \gamma_x) \right) \pm
   \sqrt{ \frac{W k_d d}{L n_c^2}  + 
     \frac{1}{4}(k_p - k_x + i(\gamma_x - \gamma_p))^2)}
     \label{eqSol1}
\end{equation}
If the radicand is positive, these are the solutions for the two polariton branches. Commonly one sets $V^2 = \frac{W k_d d}{L n_c^2}$, evaluates the system on resonance ($k_p = k_x $) and compares the terms of the radicand. Then $ 2 V < |\gamma_x - \gamma_p|$ marks the weak coupling regime, in which no mode splitting occurs and $ 2 V > |\gamma_x - \gamma_p|$ falls into the strong coupling regime with finite normal mode splitting. Note that the Rabi splitting $\Omega = 2 \sqrt{ V^2  - 
     \frac{1}{4}(\gamma_x - \gamma_p)^2)}$ only follows $V$ if the linewidths can be neglected in the radicand. As well note that a system given by two coupled, damped oscillators as:
    \begin{equation}
(k-k_p+i\gamma_p)(k-k_x+i\gamma_x) = V^2
\label{S18}
\end{equation} 
     has solutions of the form Eq. \ref{eqSol1} with $V$ as given above.
     
\begin{figure}[h!]
\centering
\includegraphics[width = 400pt]{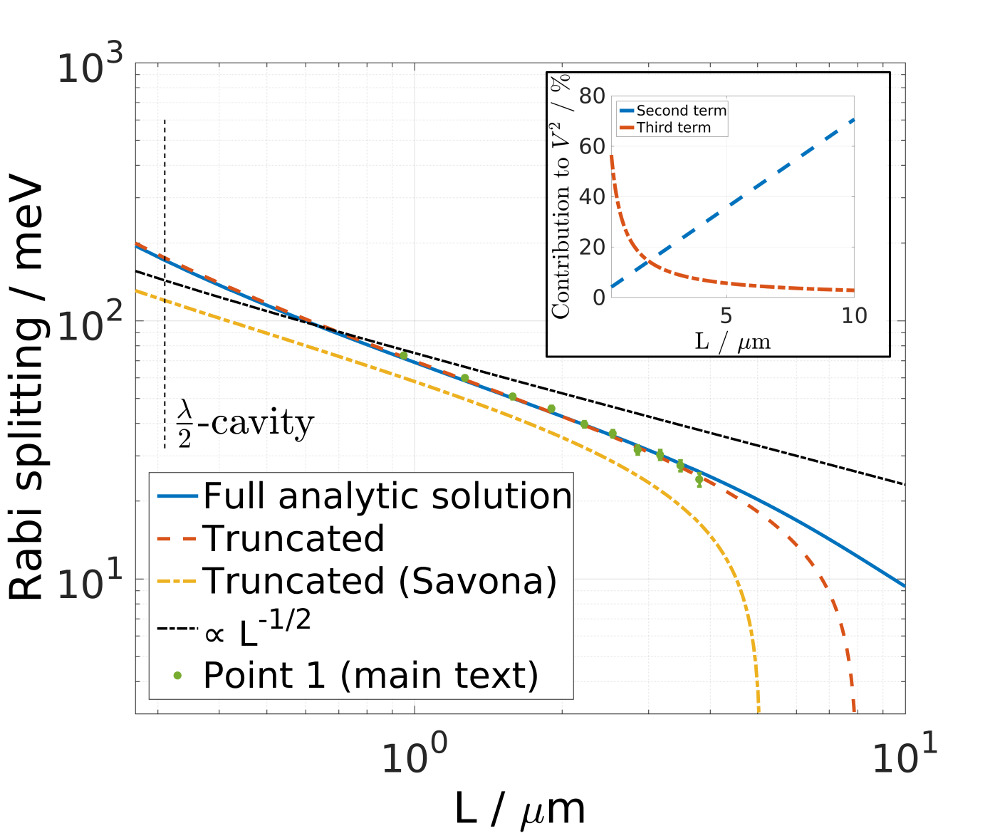}
\caption[flushleft]{\textbf{Rabi splitting $\Omega$ as obtained by different analytical expressions.} The black dashed line shows $\Omega \propto \frac{1}{\sqrt{L}}$, the blue continuous line depicts the analytic solution without truncation of terms in higher order of $d$, the red dashed line corresponds to our truncated solutions from Eq. \ref{eqTrunc}, the yellow dashed line shows the solution from Savona et al. \cite{savona_quantum_1995} (Eq. \ref{eqTruncSav}). The green dots are experimental values as presented in the main text. Parameters have the numeric value: $\epsilon_B = 20$, $\gamma_x =$ \SI{28}{\milli\electronvolt}, $R$ = 0.95, $k_x =$ \SI{2}{\electronvolt}, $k_p =$ \SI{2}{\electronvolt}, $d =$ \SI{0.8}{\nano\meter}, $n_c = 1$, $W = 21\gamma_x$ (see \cite{li_measurement_2014} for choice of $\epsilon_B  \, \& \, W$). \label{figS2}} 
\end{figure}

Now the question is, what shape the artificially introduced cavity linewidth $\gamma_p$ has in terms of the mirror reflectivity. To this end we set $r_1 = \sqrt{R}$ and follow the above procedure again. We find lengthy solutions for $k_{1,2}$ in which we can identify
\begin{equation}
\begin{split}
\gamma_p & = \frac{1-\sqrt{R}}{\sqrt{R}} \frac{1}{2 n_c L}  \\ \
V^2 & =  \frac{k_d d W}{4 n_c^2 L } \left( \frac{3 + \sqrt{R}}{\sqrt{R}}  + 2 L n_c \gamma_x \right) + \left(\frac{k_d d \epsilon_B}{2L n_c^2 \sqrt{R}}\right)^2 \\& =  \frac{1 + \sqrt{R}}{\sqrt{R}} \frac{k_d d W}{2 n_c^2 L } + \frac{k_d d W}{2 n_c} \left(\gamma_p + \gamma_x \right) + \left(\frac{k_d d \epsilon_B}{2L n_c^2 \sqrt{R}}\right)^2
\end{split} \label{eqTrunc}
\end{equation} 
In this we have neglected terms of higher order in $d$, except for one term proportional to $\epsilon_B^2$, and assumed $k_d d \epsilon_B \ll n_c$. Comparing these solutions with expressions found by Savona et al. \cite{savona_quantum_1995}
\begin{equation}
\gamma_{p_{\rm{Sav}}} = \frac{1-\sqrt{R}}{\sqrt{R}} \frac{1}{n_c L} \qquad \& \qquad V_{\rm{Sav}}^2 = \frac{1 + \sqrt{R}}{\sqrt{R}} \frac{\Gamma_0}{n_c L}  \label{eqTruncSav}
\end{equation}  
 we identify a few differences. Here $\Gamma_0$ is proportional to the oscillator strength of the quantum well. $\gamma_p$ matches the published one, the difference in the denominator stems from the different cavity geometry. The published form of $V^2$ lacks the two last terms in our truncated solution \cite{kavokin_microcavities_2011,savona_quantum_1995}. The first correction term $\frac{k_d d W}{2 n_c} \left(\gamma_p + \gamma_x \right) $is small for small linewidths $\gamma_p + \gamma_x \ll \frac{2 c}{n_c L}$, but for room temperature polariton applications it becomes sizable. In our case it corrects the value for $V^2$ by $7\%$ for $L = 1$~\SI{}{\micro\meter} and by $25\%$ for $L = 4$~\SI{}{\micro\meter}. The second term $\left(\frac{k_d d \epsilon_B}{2L n_c^2 \sqrt{R}}\right)^2$ is small for large cavity lengths and low refractive indices of the thin dispersive medium, i.e. for $\frac{k_d d \epsilon_B^2}{4 L n_c^2} \ll W$. In our case, where $\epsilon_B = 21$ is large \cite{li_measurement_2014}, it adds about $28\%$ for $L = 1$~\SI{}{\micro\meter} and $7\%$ for $L = 4$~\SI{}{\micro\meter} to the value of $V^2$ (see Fig. \ref{figS2} inset). Fig. \ref{figS2} shows the difference of these expressions. The black dashed line shows $\Omega \propto \frac{1}{\sqrt{L}}$, the blue continuous line depicts the analytic solution without truncation of terms in higher order of $d$, the red dashed line corresponds to our truncated solutions from Eq. \ref{eqTrunc} and the yellow dashed line shows the solution from Savona et al. \cite{savona_quantum_1995} (Eq. \ref{eqTruncSav}). Parameters are chosen as for the fit to our data. In particular we chose $W = 21\gamma_x = $ \SI{588}{\milli\electronvolt}. 
We have included the experimental data from one position from the main text to demonstrate the difference in slope and magnitude to the solution given by Eq. \ref{eqTruncSav}. Note that our model predicts a larger coupling for small cavity lengths, in particular we estimate a Rabi splitting of $\approx 180$~\SI{}{\milli\electronvolt} for a $L = \frac{\lambda}{2} = 310$~\SI{}{\nano\meter} cavity.
 
\vspace{100pt}

\section{Equivalence between phenomenological Hamiltonian and TMM approach for description of polariton dispersion}
\setcounter{equation}{20}
In this work we have used two different descriptions for the strongly coupled system of a cavity mode and the monolayer of WS$_2$. Here we show the equivalence between both descriptions.

In the main text we have limited our description to a single cavity mode with energy $E_c$ and a single exciton state $E_x$, coupled by a phenomenologically introduced interaction with strength $\frac{\hbar \Omega_{\rm{Rabi}}}{2}$. The system given by Eq. (1) is solved by using the Hopfield transformation resulting in mixed cavity-exciton states with lower and upper polariton operators. Eq. (2) describes the same system in a basis representation, here in the basis of cavity mode $\ket{\alpha}$ and exciton state $\ket{\beta}$. The eigenvalues of this matrix are given by:
\begin{equation}
E_{1,2} = \frac{1}{2} (E_{\rm{cav}}+E_{\rm{exc}})+\frac{1}{2}\sqrt{(E_{\rm{cav}}-E_{\rm{exc}})^2+\Omega_{\rm{Rabi}}^2}
\end{equation}
which is equivalent to Eq. \ref{eqSol1} when neglecting the linewidths of both systems. Indeed the relation $\Omega_{\rm{Rabi}} = 2V$ is motivated by this equivalence. We thus see, that both the phenomenological Hamiltonian in Eq. (1) and the TMM approach, which is based on classical electromagnetism and a simple oscillator model, results in the same dispersion. Other parameters, such as the polariton linewidths, are only explained by the classical model. For a more extensive description of strongly coupled systems we point the reader to the literature: Kavokin and coworkers give a comprehensive review of polariton formation within the framework of three different formalisms: the classical, the semiclassical and the full quantum description \cite{kavokin_microcavities_2011}.
Savona et. al. have described the polariton formation in quantum well microcavities with both a semiclassical TMM approach and a quantum theoretical approach \cite{savona_quantum_1995}. Equations from both approaches (Eq. (7) and (10) in Ref. \cite{savona_quantum_1995}) give the same result which we have found in Eq. \ref{S18}. While Savona et al. start off by supposing quantised radiation modes in their Hamiltonian Eq. (8) in Ref. \cite{savona_quantum_1995}, a more rigorous derivation entailing quantisation of both the photon field and the exciton state is given by Gerace et al. \cite{gerace_quantum_2007}.

\putbib
 
\end{bibunit}

\end{document}